# Reconfigurable Hardware Accelerators: Opportunities, Trends, and Challenges

Chao Wang, School of Computer Science, University of Science and Technology of China. Hefei Anhui 230027
Wenqi Lou, School of Computer Science, University of Science and Technology of China. Hefei Anhui 230027
Lei Gong, School of Computer Science, University of Science and Technology of China. Hefei Anhui 230027
Lihui Jin, School of Computer Science, University of Science and Technology of China. Hefei Anhui 230027
Luchao Tan, School of Computer Science, University of Science and Technology of China. Hefei Anhui 230027
Yahui Hu, School of Computer Science, University of Science and Technology of China. Hefei Anhui 230027
Xi Li, School of Computer Science, University of Science and Technology of China. Hefei Anhui 230027
Xuehai Zhou, School of Computer Science, University of Science and Technology of China. Hefei Anhui 230027

With the emerging big data applications of Machine Learning, Speech Recognition, Artificial Intelligence, and DNA Sequencing in recent years, computer architecture research communities are facing the explosive scale of various data explosion. To achieve high efficiency of data-intensive computing, studies of heterogeneous accelerators which focus on latest applications, have become a hot issue in computer architecture domain. At present, the implementation of heterogeneous accelerators mainly relies on heterogeneous computing units such as Application-specific Integrated Circuit (ASIC), Graphics Processing Unit (GPU), and Field Programmable Gate Array (FPGA). Among the typical heterogeneous architectures above, FPGA-based reconfigurable accelerators have two merits as follows: First, FPGA architecture contains a large number of reconfigurable circuits, which satisfy requirements of high performance and low power consumption when specific applications are running. Thus, it can achieve a high energy efficiency rate. Second, since latest applications feature various types and rapid iterations, it takes an extended period of adopting ASIC to design accelerators. By contrast, the reconfigurable architectures of employing FPGA performs prototype systems rapidly and features excellent customizability and reconfigurability. Nowadays, in top-tier conferences of computer architecture, emerging a batch of accelerating works based on FPGA or other reconfigurable architectures. Meanwhile, FPGA-based accelerator platforms have become one of most critical technologies in the industry. Internationally known companies, such as Intel, Microsoft, and so on, take reconfigurable platforms as an important measure to construct next-generation heterogeneous accelerators which are widely applied to data centers and embedded devices. To better review the related work of reconfigurable computing accelerators recently, this survey reserves latest high-level research products of reconfigurable accelerator architectures and algorithm applications as the basis. In this survey, we compare hot research issues and concern domains, furthermore, analyze and illuminate advantages, disadvantages, and challenges of reconfigurable accelerators. Since Computer Architecture is one of the hugest gaps of development in computer domains, we take the reconfigurable accelerators orientation as a starting point and focus on analyzing opportunities and challenges for domestic architecture researchers, which are caused by the development technology of FPGA and reconfigurable accelerators. In the end, we prospect the development tendency of accelerator architectures in the future. Overall, this survey aims to provide a reference for computer architecture researchers.

CCS Concepts: •

**KEYWORDS**

# 1 INTRODUCTION

With the emerging big data applications of artificial intelligence, bioinformatics, and others, the compute-intensive features and memory-intensive features of the applications put forward increasingly requirements on the computing ability of hardware platforms. However, the reality is that since the design of general-purpose CPUs needs to tradeoff performance, power consumption, size, versatility, and many other factors, its computation power is difficult to meet the needs of such emerging applications. At present, the architecture solutions commonly used in the big data area are built on cloud and General Purpose Graphics Processor Unit (GPGPU) computing platforms. The cloud computing platform is usually composed of a large number of CPU-based homogeneous server nodes, in which multiple nodes cooperate with each other and work together. Although the cloud computing platform can achieve a preferable acceleration effect, its internal single node computing efficiency is still low. Furthermore, the overall performance is constrained by the network bandwidth, and hardware costs and energy consumption are also enormous. GPGPU has a large number of hardware-level threads and parallel processing units, the past mainly used in the field of graphics processing with high computational parallelism. With the advent of general-purpose computing frameworks such as Compute Unified Device Architecture (CUDA) and son on, GPGPU is also widely applied in the field of general-purpose computing and has significant performance improvements over general-purpose processors. However, behind the good acceleration, tremendous energy expense is also an issue that such platforms cannot avoid.

To improve the computing power while reducing energy consumption and improving computing efficiency as much as possible, researchers begin to utilize dedicated hardware accelerators to handle these applications demanding a higher computing and memory access. The current design of such hardware accelerators is based primarily on Application-specific Integrated Circuit (ASIC) and Field Programmable Gate Array (FPGA). ASIC is a type of integrated circuit chip designed and developed for the specific purpose, with high-performance, low power consumption, small size and so on. However, the ASIC design cycle is long, and the mask cost is high, and the mask cost increases exponentially as the line width decreases, which further increases the design cost of the ASIC. On the contrary, FPGA as a representative of reconfigurable devices, combining the flexibility of software computing with the efficiency of hardware computing, makes the design cost of FPGA-based hardware accelerators far lower than that of the ASIC-based hardware accelerators. Also, FPGA has a more extensive range of applications due to its customizability. Furthermore, the advent of high-level synthesis tools for FPGAs in recent years has drastically reduced the R & D cycle and design threshold, bringing great convenience to FPGA design.

Due to reconfigurable devices such as FPGA possessing lots of superiority in the design of accelerators, the related work has drawn full attention from industry and academia. In the industrial field, Intel, Nvidia, Google, Microsoft, ARM, Apple and other well-known companies have invested heavily in the development of reconfigurable computing platforms. For instance, Apple Inc. of the United States pioneers the process of embedding reconfigurable coprocessors into handheld embedded devices by embedding an FPGA chip in an iPhone 7 handset. In the academic field, at the recent top-tier international conferences on architecture, a batch of reconfigurable architecture research results and the research results of the heterogeneous accelerator based on it emerge. Such accelerators efficiently increase computing speed and lessen energy consumption when dealing with applications in specific fields (such as neural networks, graph computation, and data mining). Relevant work includes a series of research achievements of neural network accelerator proposed by Nurvitadhi research group, as well as DianNao series essays introduced by CAS.

Overall, in recent years, accelerator design based on reconfigurable devices has not only achieved a higher degree of development, in the field of hardware architecture, also made impressive strides in software applications and algorithms. Enhance the security and reliability of the reconfigurable architecture, which provides a solid foundation for building a new type computer system and promoting the industrialization of artificial intelligence chips and systems. Whereas, we also need to recognize that there are still many problems in the current architecture, such as the insufficiency of parallelism, security, and reliability and high energy consumption, which bring new challenges and opportunities to the computer industry. Based on the recent high-level scientific research achievements related to reconfigurable computing accelerator architecture and application of algorithms, this survey will objectively summarize relevant research hot spots, analyze and demonstrate the advantages, disadvantages, and challenges of reconfigurable computing accelerators in detail. This survey focuses on the challenges and opportunities of the reconfigurable computing accelerator architecture in such hot topics as neural networks, graph computing, and database, bioinformatics. In the end, we outlook the development trends and prospects of reconfigurable computing accelerator architecture, hoping to provide a reference for the related research in the field of computer architecture in our country.

## 2 The State-of-the-arts

This chapter focuses on leading research on reconfigurable computing accelerators and expounds the current research status in the field of reconfigurable computing accelerators in both industry and academia, mainly covering FPGA-based accelerator architectures and application-specific solutions.

### 2.1 FPGA-based accelerator architecture research

In recent years, with the rise of big data applications, the demand for handling explosive data continues to grow. Meanwhile, Moore's Law of CPU has entered its dusk. Apparently, it has been difficult to fit the scale of exponential growth of machine learning and web services. All the time, people accelerate common computing tasks by using custom hardware. However, fast-changing applications and algorithmic iterations require that this customized hardware can be reprogrammed to enable it to accelerate new types of computational tasks. Field Programmable Gate Array (FPGA) is precisely a kind of reconfigurable hardware, and its interior contains a large number of reconfigurable circuits, with good customization and scalability. Moreover, it could comply with the high performance and low power operational requirements of a given application, resulting in a high-performance ratio (performance to power ratio). Compared to CPU and GPU, FPGA is inherently an instructionless, shared-memory-less architecture that better accelerates algorithms in the big data field that includes many compute-intensive and communications-intensive tasks. Besides, using FPGA in the field of big data can customize computing unit density and balance tuning chip area, power consumption, performance and other indicators [1]. Therefore, FPGA-based accelerator architecture research has become one of the principal directions in the current academic and industrial research to optimize acceleration.

Over the past decade, Intel, Microsoft, Google and other famous enterprises and many institutions of higher learning have shown great interest in this area. They invest continually and vigorously in FPGA-based Accelerator architecture research and launch a large number of high-performance FPGA accelerator architectures.

As early as 2014, some well-known foreign universities and research institutes hatch numerous enterprises (such as Maxeler, BlueBee, Convey Computer, and Nallatech) that use FPGA to provide high-performance solutions for big data applications. They primarily offer FPGA-based acceleration solutions for big data areas such as financial analysis, social media, and scientific computing, demonstrating the practical use and bright future of FPGA in the big data field and laying a good foundation for the development of FPGA-based Accelerator architecture.

After 2015, with the losing efficacy of Moore's Law and the rise of heterogeneous multi-core platform research, many domestic and foreign enterprises turn their attention to FPGA. In June 2015, Intel acquires FPGA maker Altera and proposes the HARP project [2], which aims to use Xeon CPUs and FPGAs to build a teaching-related platform that faces new-type algorithms and applies to the programming system, architecture design, and algorithm research. In the same year in November, IBM releases SuperVessel [3], a GPU-based and FPGA-based cloud architecture oriented to the OpenPOWER cloud platform. After that, IBM successively releases the services of Hyper Cloud, Big Data, and FPGA Accelerator to provide developers with a platform for creating and testing mobile services and data analysis and other emerging areas. In February 2016, Amazon launches Amazon EC2 F1 [4], an FPGA computing instance, on its cloud service, AWS, which makes it easier for users to deploy FPGA accelerators to address the complex issues in science, engineering, and business that require high bandwidth, enhanced networking, and ultra-high computing power, especially for time-sensitive applications such as clinical genes, financial analysis, and machine learning. Soon after, foreign startup Teradeep introduces an FPGA-based architecture for accelerating image recognition, which has a suitable applied value in image recognition and video analysis.

In September 2016, Wired magazine publishes an article "Microsoft Bets Its Future on a Reprogrammable Computer Chip" and describes the past and present progress of the CataPult project [5]. As early as 2011, to meet new, super large scale application requirements, Microsoft launches the CataPult project, which aims to accelerate and optimize AI services and applications by utilizing FPGAs. Microsoft deployed FPGA roughly experiences a dedicated FPGA cluster, dedicated network connection, and shared server network three stages. In the first stage, similar to the supercomputer, Microsoft adopts a dedicated cluster to deploy FPGAs. In other words, like a supercomputer composed of FPGAs, the dedicated cluster is entirely populated with FPGA accelerator cards. Microsoft initially utilizes the BFB Experimenter Board and deploys six FPGAs on one PCIe card, four PCIe cards on each 1U server. In this case, 24 FPGAs are deployed on each 1U server, which triggers the FPGA of different machines difficult to communicate, network delay difficult to stabilize, FPGA-specific cabinet single point of failure, server customization, and other issues [6]. In the second stage, to ensure server isomorphism in the data center, Microsoft assigns an FPGA on each machine, using a dedicated network connection between FPGAs [7]. This method successfully accelerates Microsoft Bing search engine through FPGA and achieves good results. In the third stage, Microsoft deploys FPGAs between network adapters and switches to accelerate network functions and storage virtualization. At the same time, Microsoft applies this solution to its cloud server Azure and thereby expands the connections between FPGAs to the size of the entire data center. At this point, the high-bandwidth, low-latency FPGA network interconnect layer acts as an acceleration medium for the network switching layer and traditional server software. Moreover, Microsoft puts forward the concept of Hardware as a Service [8] (HaaS), namely regarding hardware as a kind of cloud computing service. This idea has much promoted the utilization and large-scale deployment of FPGA.

In November 2016, Inspur and Intel jointly release the F10A, an FPGA accelerator card jointly developed by both parties, which meets OpenCL's high-density, high-performance and high-bandwidth needs and lays a certain hardware foundation for FPGA accelerator architecture research conducted by domestic enterprises and universities. In January 2017, Tencent introduces the first high-performance heterogeneous computing infrastructure in China, FPGA Cloud Services, to address the issue of efficiently processing images for its social software due to the rapid growth of the mobile Internet. Meanwhile, it provides deep learning model services to other developers, which directly drives the research and investment of other domestic companies in this area. In the same month, Ali cloud also releases the FPGA cloud program, revealing its determination in this area. In June 2017, Baidu releases its FPGA Cloud Server based on previous research on FPGA acceleration, mainly oriented to applications such as Convolutional Neural Network (CNN) image classification, RSA decryption acceleration, and gene sequence alignment. Besides, Google, Facebook, and other companies have also researched this area and launched projects related to FPGA accelerator.

In the academic field, researchers in the United States, China, Italy, Britain, Spain and other countries also conduct extensive research on heterogeneous reconfigurable systems. Such as Berkeley's RAMP [9], Xilinx's Zynq-7000 [10] and Cornell's ReMAP [11], ZCluster [12] at the University of Toronto, MOLEN [13] of TUDelft University, Multicube [14] in Europe, FPMR [15] at Tsinghua University, RCC [16] at Jiangnan Institute and SOMP [17] at USTC are all typical research projects in the field of heterogeneous reconfiguration.

RAMP [9] is a reconfigurable system-on-chip architecture designed for on-chip multicore processor research, related operating systems, and programming model development. It mainly implements a shared memory multi-core system, that is, all cores share an off-chip DRAM memory. The project team utilizes the Aparc architecture as a reconfigurable multi-core processor architecture and deploys it on Xilinx's XUP and BEE3 development boards. The Zynq-7000 [10], a programmable heterogeneous multi-core system-on-chip introduced by Xilinx Inc., consists of programmable logic devices and processing systems and features configurability, flexibility as well as high performance. Among them, programmable logic devices are mainly Xilinx 7 series FPGA which can be served as a dedicated processor to accelerate a specific task. The core of the processing system is an integrated ARM processor, generally used for routine tasks, and can operate independently out of programmable logic. ReMAP [11] is a heterogeneous multi-core architecture that is primarily suitable for heterogeneous multi-core system-on-a-chip that integrates programmable logic resources. ReMAP contains several general-purpose processor clusters and dedicated programmable logic clusters. ReMAP uses general-purpose processors for general tasks and deploys threads with specific tasks into the corresponding dedicated programmable logic clusters, which effectively speeds up task execution and improves system performance. Zcluster [12] is a cluster platform based on the Zedboard development board, utilizing the Hadoop programming model for computing node management. Zcluster mainly consists of eight Zenboard development boards, in which each development board integrates a zc702 reconfigurable chip. Concretely, Zcluster implements a standard Fourier filter and performs 3.3 times faster than a single ARM processor and 1.2 times faster than an 8-node ARM cluster without reconfigurable logic hardware acceleration. MOLEN [13] is a compiler for reconfigurable processors, the primary function of which is to automatically compile the C language to binary code available for reconfigurable platforms such as FPGAs. On FPGA-based PowerPC processors, MOLEN implements code generation, allocation of register and stack, enabling the expanded PowerPC back-end to generate instructions for reconfigurable hardware platforms. Multicube [14] is a cluster platform composed of 8 development boards. Its primary purpose is to research the on-chip multi-objective design space search algorithm. Sixty-four FPGA devices (64 FPGA devices organized in a matrix of 8 × 8) are integrated on each board, resulting in an 8 × 8 × 8 cube structure. FPGAs in the Multicode take pulse bus connection, and each FPGA device accepts the data input from the previous FPGA and outputs the result of the calculation to the next FPGA device. Experiments show the execution speed 359 times faster than that of 2.5GHz Intel Quad-Core Xeon processor. Maxwell [18] proposes an FPGA-based general-purpose supercomputer, Maxwell, which is composed of 64 FPGAs and can be applied in three complex application scenarios (petroleum, finance, and medicine).

FPMR [15] proposed by Tsinghua University is an FPGA-based MapReduce framework that encapsulates functions such as task management, communication, and synchronization to facilitate the design of hardware. It has a good acceleration effect on some algorithms such as RankBoost, SVM, and PageRank. Experimental results demonstrate that compared with utter hand-designed version, the performance is 33.5 times higher. SOMP [17] is a service-oriented heterogeneous multi-core reconfigurable on-chip computing system on a single FPGA, the framework of which introduces service-oriented concept and service body-execution flow model into the design of heterogeneous multi-core reconfigurable on-chip systems. On the one hand, the architecture integrates diverse hardware computing resources so that it can take full advantage of the strengths of multi-processor systems on chip (MPSoC) to obtain high computing performance; On the other hand, employing the reconfigurable technology, this framework can restructure the hardware resources. Adjusting the corresponding software and hardware task allocation and scheduling strategy simultaneously, SOMP further improves the throughput and task parallelism of the platform. Based on the SOMP architecture, the research group also proposes SoSoC [19], a service-oriented system-on-chip framework that integrates both embedded processors and software-defined hardware accelerators as computing services on a single chip. Benefiting from the high computing performance of MPSoC and the flexibility of SOA, SoSoC provides well-defined programming interfaces for programmers to utilize diverse computing resources efficiently. Experimental results demonstrate that the service componentization over original version is less than 3%, while the speedup for typical software Benchmarks is up to 372x. Although MPSoC can provide increasingly speedups to diverse applications, it's still limited by the interconnection and communication through processors. However, traditional interconnect architectures cannot obtain both performance and flexibility. To solve this problem, Wang et al. [20] propose a flexible high-speed star network based on peer to peer links on FPGA. By utilizing fast simplex links (FSL) high-speed bus channels and programming interfaces for demonstration, this method can achieve both high efficiency and flexibility. SODA [21] is a software-defined FPGA based accelerator for big data. It could reconstruct and reorganize the acceleration engines according to the requirement of the various data-intensive applications.

Regarding supporting reconfigurable technologies, due to upper-application software developers lacking knowledge of reconfigurable hardware details, programming wall problems in reconfigurable systems persist. Thus, research at the operating system level has focused primarily on providing universal programming interfaces and programming models for underlying reconfigurable hardware logic. David Andrews et al. propose a unified multi-threading model Hthread [22] for reconfigurable systems. The threads running on the general-purpose processor and on the reconfigurable resources are respectively defined as software threads and hardware threads. The operating system takes unified management of both threads, thus shielding the differences in computing resources and lowering the difficulty of programming. Furthermore, there are also some representative operating systems such as BORPH [23], ReconOS [24] and OS4RS [25]. Among them, UC Berkeley's BORPH [23] is a UNIX-like reconfigurable operating system that holds the configurable logic configuration information in an executable file and implements inter-thread communication through FIFO buffering. Since BORPH is only responsible for the configuration of the bit streams and cannot schedule reconfigurable resources, support for the dynamic reconfiguration of the platform remains to be explored further. ReconOS [24] appends the function of multithreaded programming of reconfigurable devices on existing embedded operating systems. ReconOS provides CPU-FPGA systems with interfaces to create, communicate and synchronize hardware and software threads, the main work of which is to provide a POSIX interface to standard operating systems. The main features of ReconOS are that it provides the same control interface for the execution of hardware threads as it provides for software threads. The hardware and software threads have the same packaging interface, which can realize a seamless connection between each other's call and communication. The advantage is that the designer does not need to consider the thread is a software thread or hardware thread, which effectively reduces the design difficulty. OS4RS [25] is a system developed for reconfigurable, the main feature of which is that it packages the tasks of hard and software uniformly by utilizing hardware abstraction layer. OS4RS customizes interfaces for execution units of hardware and makes hardware execution devices integrate the ability of communication between threads. Moreover, OS4RS provides thread communication interfaces that allow software and hardware interfaces to communicate with each other. David Koeplinger and others from Stanford University [26] use parametric templates to represent hardware and design a framework by utilizing techniques such as template-level models and design-level artificial neural networks to interpret hardware layout tools. This framework solves the current FPGA-oriented tools failed to support advanced programming, resources estimation, and rapid automatic design of space exploration and other issues. Christoforos Kachris and Dimitrios Soudris investigate FPGA accelerator frameworks that used in the data center [27] and describe relevant usage in cloud computing applications such as FPGA accelerators related to MapReduce and their qualitative classification and comparison. Dionysios Diamantopoulos and Christoforos Kachris propose the reconfigurable MapReduce accelerator HLSMapReduceFlow [28] based on FPGA, which can accelerate the processing of map computing kernel. It allows the data center to seamlessly develop FPGA-based hardware accelerators by extending the current High-Level Synthesis (HLS) tool flow to include the MapReduce framework. Zeke Wang et al. design and complete a MapReduce framework for FPGA, Melia [29], which has certain advantages over CPU and GPU. His another paper [30] presents an FPGA-based performance analysis framework that could reveal performance bottlenecks and thereby guide code tuning of OpenCL applications on FPGAs. For privacy issues of FPGA cloud services, Lei Xu, Weidong Shi, and Taeweon Suh propose FPGA cloud – PFC [31] used for privacy protection calculation in public cloud environments. As an important use of cloud computing, it applies PFC to the MapReduce programming model and extends the FPGA-based MapReduce pipelines with privacy protection. At the same time, it also employs proxy re-encryption to support dynamic allocation of trusted FPGA devices as mappers and restorers. Muhsen Owaidad et al. propose a framework, Centaur [32] running on an FPGA. The framework allows to dynamically allocate FPGA operators and query plans, thus calling these operators when needed, and run them mixed on CPUs and FPGAs. Meanwhile, it provides a realistic solution for accelerating SQL that is compatible with existing database schemas, which makes it possible to explore FPGA-based data processing further. Focused on developing infrastructures, Jason Cong et al. enable FPGA-based acceleration in the data center [33]. They provide an initial version of an integrated solution that includes FPGA-based accelerator generation, the automatic compilation of runtime accelerator resource scheduling and management, and an accelerated library of custom calculations for big data applications. Overall, FPGA-based reconfigurable computing accelerators have achieved academia leading results in architecture, operating systems, programming tools, and so on.

## 2.2 Accelerator for specific applications

### 2.2.1 Neural Network Accelerators

In the new type of big data and artificial intelligence applications in recent years, neural networks have become a hot research area and also become an important application driver for the development of new architectures. With an increasing demand for accuracy, power consumption and computation time, a growing number of scholars are keen on designing and implementing accelerators suitable for neural networks.

ASIC, as an application-specific integrated circuit, has high execution efficiency for specific applications and algorithms. Therefore, ASIC-based neural network accelerator research is pervasive. For instance, Cnvlutin [34] proposes a deep convolutional neural network that could remove invalid neurons, which could reduce the use of computing resources and accelerate memory access and in turn optimize performance and energy consumption. Eyeriss [35] presents a low-power data flow structure for convolution neural networks and replaces GPU's accelerators of the SIMD / SIMT architecture with the CNN accelerators that adopt a new type data movement model RS. Neurocube [36] proposes a programmable neural network acceleration architecture based on 3D stacked storage. It utilizes particular logic module settings to accelerate neural network computation while enjoying the storage of enormous internal bandwidth and eliminating unnecessary and expensive data movement. Minerva [37] is a type of low-power, high-precision deep neural network accelerator that proposes an accelerator design and optimization process based on design space search. RedEye [38] proposes an image sensor structure specific to the simulated convolutional neural network of moving vision. It employs a modular column-parallel design approach to lessen the complexity of analog design, thereby facilitating the reuse of physical design and algorithms. In allusion to the algorithms of the neural network, DianNao [39] designs an accelerator with a high throughput, capable of performing 452 GOP/s in a small footprint of 3.02mm2 and 485mW. Compared to a 128-bit 2GHz SIMD processor, the accelerator is 117.87x faster, and it can reduce the total energy by 21.08x. DaDianNao [40] proposes a machine-learning supercomputer, which is a SIMD implementation based on DianNao. The difference is that the weight matrix used for computation is solidified into the local eDRAM, reducing the number of memory reads. It achieves a speedup of 450.65x over a GPU and reduces the energy by 150.31x on average for a 64-chip system. PuDianNao [41] presents a polyvalent machine learning accelerator that provides acceleration for seven common machine learning algorithms with 1.2 times faster acceleration and 128.41 times lower power consumption than the NVIDIA K20M GPU. Cambricon [42] proposes a neural network instruction set that directly deals with large-scale neurons and synapses, in which with a single instruction, a group of neurons can be processed. Moreover, the instruction set also provides a series of specialized support for the on-chip transmission of neurons and synapse data. Cambricon-X [43] is a deep learning accelerator that can efficiently deal with sparse networks. The accelerator filters out the zero-valued neurons by marking the non-zero neurons one by one before sending the neurons into the computing unit. Then, the weight buffer inside the calculation unit only stores the weight data corresponding to non-zero neurons, thus eliminating the redundant calculation and weight storage. EIE [44] proposes an efficient inference engine for compressed deep neural networks and gives the hardware implementation of the CNN after processed by the compression algorithm. This hardware implementation could distribute the compressed neural network parameters to the SRAM entirely, significantly reducing the number of DRAM access, while DRAM access is the most energy-intensive operation of the traditional CNN accelerator. The above work sufficiently shows that the ASIC-based neural network accelerator has become a hotspot of the current research in the field of architecture.

In addition to ASICs, constructing heterogeneous accelerators based on GPUs is a widely adopted solution for setting up rapid prototype system. Reference [45] studies the memory efficiency of different layers of CNN and reveals the impact of data layout and memory access mode on performance. Reference [46] puts forward a hardware mechanism with low delay and high efficiency to skip the multiplication operations that the input is zero, and also proposes the data reuse optimization of the addition operation. This method improves the utilization of local registers and reduces on-chip cache access. NVIDIA [47] presents a run-time memory manager for neural network applications, which enables deep neural networks to be mapped onto CPU memory and GPU onboard memory while being trained and maximizes training speed. Reference [48] proposes an efficient implementation of the large-scale recurrent neural network on GPU and proves the scalability of this implementation on the GPU. It exploits the potential parallelism of recurrent neural networks and proposes a fine-grained two-stage pipeline implementation. Reference [49] conducts a comprehensive comparison of a series of CNN open source implementations based on GPUs through a wide range of parameter configurations, investigates potential performance bottlenecks, and points out further optimization methods. In reference [50], two scheduling strategies are proposed to optimize the target detection tasks and have improved the performance of the system. Overall, the heterogeneous acceleration system based on GPU is relatively mature at present and can rapidly iterate between GPU board and neural network topology models and algorithms.

In the design of FPGA-based heterogeneous accelerators, ESE [51] proposes an efficient speech recognition engine for sparse LSTM, which not only lowers the algorithm to a smaller size also supports the compressed depth learning algorithm on the hardware level. Nurvitadhi et al. [52] assess the gap of performance and energy efficiency between FPGAs and GPUs in accelerating next-generation deep neural networks and conclude that as hardware resources increase, FPGAs will become the choice for the next-generation DNN acceleration platform. Nurvitadhi et al. [53] have also studied variants of circular neural networks and propose storage optimization to avoid vector multiplication of partially dense matrices. DNNWeaver [54] is a framework that can automatically generate synthesizable FPGA accelerators according to the high-level neural network models. Escher [55] is an FPGA-based CNN accelerator, which is equipped with a flexible data buffer to ensure the balance of input and weight conversion bandwidth. It could effectively reduce the overall bandwidth requirement. Reference [56] presents a Fused-Layer, which focuses on the problem of data flow in the convolutional layer and reduces the transfer of off-chip feature mapping data by 95%. Reference [57] uses Roofline model analysis design to optimize FPGA-based CNN accelerators. Reference [58] mainly solves the problem of low dynamic resource utilization rate of CNN accelerator based on FPGA and designs a new type of CNN accelerator, which leads to significant performance improvement.

Focused on the implementation of the prediction process, data access optimization, and pipeline structure in large-scale deep learning neural networks, reference [59] introduces a deep learning prediction process accelerator based on FPGA.Compared with Core 2 CPU 2.3GHz, this accelerator can achieve a promising result. Reference [60] proposes a uniformed representation of neural networks applicable to convolutional layers and fully connected layers and designs software and hardware co-calculation engine - Caffeine, which combines Caffe, the industry-standard software deep learning framework. Compared with the traditional CPU and GPU, it has considerable performance and energy efficiency improvements. When using FPGA to accelerate the feedforward calculation process of convolution neural networks, due to different network topologies and various FPGA hardware resources, the accelerator has a relatively serious waste problem in resources of computing and memory access. To simplify the design process, reference [61] proposes DeepBurning, a set of the FPGA-based neural network accelerator development framework. By analyzing common neural network topologies, authors summarize the frequently-used components in a series of networks and form a component library through RTL level description. Users only need to provide the upper layer description of the network topology as well as hardware resource constraints, and the neural network integrator (NN-Gen) in the framework can automatically analyze the network characteristics. Then, it will select the appropriate components in the component library to form a hardware network, and the corresponding control flow, data flow, and data layout program, according to the hardware constraints. This framework makes the upper application designers able to utilize FPGA to accelerate neural network computing as easily as using Caffe, which greatly enhances the applicability of FPGA in this field. Reference [62] proposes to optimize the convolution cycle by quantitatively analyzing and optimizing the design objectives of the CNN accelerator based on multiple design variables. By searching the design variable configurations, it also presents a specific data flow of hardware CNN acceleration to minimize the memory access and data movement. Similarly, reference [63] introduces a performance analysis model to deeply analyze the resource requirement of CNN classifier kernels and available resources on modern FPGAs. The authors identify that the key performance bottleneck is the on-chip memory bandwidth. Therefore, they propose a new type of core design that can efficiently locate such bandwidth limitations, which can provide an optimal balance between computation, on-chip, and off-chip memory access. Reference [64] implements automatic code generation of CNNs on FPGAs, and the conclusion shows that it has better performance and saves developing round time. Reference [65] proposes an FPGA-based scalable accelerator, DLAU, which is mainly used to accelerate large-scale deep learning networks. DLAU consists of three pipeline units and enhances the locality of learning applications as well as system throughput by utilizing fragmentation technology, achieving up to 36.1x speedup compared to the Intel Core2 processors, with the power consumption at 234mW. Reference [66] proposes a high-performance convolutional neural network accelerator, which maps all the layers on one chip to improve the concurrency and achieves high system throughput and high resource utilization. Meanwhile, a batch-based computing method is applied on fully connected layers to increase the memory bandwidth utilization, resulting in that accelerators achieve a peak performance of 565.95 GOP / s and 391 FPS under 156MHz clock frequency. FlexFlow [67] is a type of flexible data flow acceleration architecture applied for CNNs. The authors note that many current CNN accelerators make use of inherent parallelism, but there is a big mismatch between the parallel types supported by computing engine and the dominant parallel types of CNN workloads. This mismatch seriously degrades resource utilization of existing accelerators. FlexFlow effectively mitigates the mismatch problem and acquires 2-10x performance speedup and 2.5-10x power efficiency improvement compared with three state-of-the-art accelerator architectures. FP-DNN [68] points out that traditional accelerator development processes make it hard for FPGA developers to keep up with the rapid update of DNNs. Therefore, it proposes an end-to-end framework that takes TensorFlow-described DNNs as input and automatically generates the hardware implementation on FPGA boards with RTL-HLS hybrid templates. Experiments show that DNNs, LSTM-RNNs and Residual Nets implemented by FP-DNN all have better performance and flexibility. Reference [69] proposes the RNNLM (Recurrent Neural Network based Language

Model) accelerator based on FPGA, the main work of which is that it improves the parallelism of the RNN training scheme and lowers the computing resource requirement for computing efficiency as well as the data load. Finally, the accelerator attains higher performance and training accuracy. Reference [70] proposes a CNN acceleration based on embedded FPGA and points out that convolutional layers are compute-intensive and fully-connected layers are storage-intensive. Therefore, dynamic-precision data quantization methods and the convolver design could be applied to all layers of the CNN to improve the bandwidth and resource utilization. Reference [71] proposes the use of heterogeneous algorithms to accelerate FPGA-based CNN accelerators. The emphasis is on applying Winograd minimum filtering theory to CNN algorithms, with the advantage of reducing the use of computing resources but placing higher demands on storage bandwidth.

Besides the above three typical accelerators, there is an RRAM-based crossbar accelerator [72] specific to the feedforward processing of binary convolution neural networks. Adopting low bit-level RRAM devices and low bit-level ADC/DAC interfaces in RRAM-based computing system enables faster reading and writing operation and higher energy efficiency. Furthermore, when the crossbar cannot store all the parameters of one layer, the accelerator designs the matrix division and pipelining implementation.

With the development of new storage technologies (such as 3D-stacking technology [73]), the research on the acceleration of neural networks based on new memory devices is also increasingly widespread. PRIME [74] proposes a ReRAM-based neural network memory computing architecture, in which ReRAM arrays can be used to augment the capacity of ordinary main memory while serving as an accelerator for neural network applications. ISAAC [75] proposes a cross-switch simulation accelerator for convolutional neural networks. To implement a pipelining-based organization to accelerate the calculation of different layers of neural networks, ISAAC takes eDRAM as a register for data between pipelining. PipeLayer [76] is a ReRAM-based PIM accelerator for CNNs that support both training and testing. RESPARC [77] is a reconfigurable and energy-efficient architecture built-on Memristive Crossbar Arrays (MCA) for deep Spiking Neural Networks (SNNs), which utilizes the energy-efficiency of MCAs for inner-product computation and realizes a hierarchical reconfigurable design to incorporate the data-flow patterns in an SNN.

### 2.2.2 Bioinformatics-Related Accelerators

Bioinformatics is an essential field of utilizing FPGA for acceleration. Researchers at UIUC have implemented accelerators that use the Pair HMM forward algorithm for DNA variation recognition [78]. This work makes the PEs connected to a ring and utilizes parallel processing nature of diagonal to schedule PE. There are mainly three optimization measures: 1) The split of calculation mode; 2) The realization of assembly line technology; 3) By reducing the size of a single PE ring and increasing the number of PE rings, the number of idle PEs is reduced efficiently. Researchers at the University of South Carolina make it more suitable for FPGA implementations by altering the NCBI BLAST algorithm [79]. The authors implement the index by transforming the database into a filter stored as a hierarchical arrangement of three tables, in which the first two layers are stored in the FPGA and the third layer in the off-chip memory. Compared with the original algorithm, this approach can efficiently reduce the number of I/O requests.

Researchers at the University of California use FPGAs to boost hardware accelerators in the Seeding stage in gene fragment matching [80]. The two main challenges for accelerating this process are 1) how to process a vast number of short sequence matching, and 2) how to hide the frequent long-term memory access when over 50% of one of the processing engines (PEs) cycle is the storage access delay. The authors propose a scalable array-based architecture, which is composed of many PEs to process large amounts of data simultaneously for the demand of high throughput.

Researchers at the University of Tsukuba propose a variable length hash method for faster short read mapping on FPGA [81]. The authors formally describe the amount of computation in the hash-index method, Tc = Cc * | SR |. Where Tc is the total computational time, and Cc is the average times of comparisons per seed and SR is a set of seeds extracted from a given short sequence. After investigation and experiment, the authors find out that in the existing hash-index method, for different seeds, the value of Cc changes much, resulting in an impact on performance. So the authors have devised a new set of hash methods that effectively reduce the Cc / Cmin value.

In 2012, researchers at CAS Institute for Computing conduct a study of short reads mapping accelerators based on FPGA [82]. The accelerator is mainly composed of the controller, PE array, buffer, data bus and DMA. Among them, accelerator interacts with the CPU by DMA to perform data interaction, and the PE only realizes the string comparison function, mainly including two shift units, two XOR gates, and 1 OR gate. From 2013 to 2014 [83-85], researchers from the University of Science and Technology of China have conducted hardware acceleration to the long-sequence aligning algorithm that uses the Smith-Waterman local alignment algorithm. Utilizing the nature of the Smith-Waterman local alignment algorithm - "the highest score always appears near the diagonal," the authors efficiently solve the problems in the long sequence aligning that the number of mismatches increases and the length that needs to be compared is too long. Furthermore, researchers at the University of Science and Technology of China also present a heterogeneous cloud framework to speed up the processing of big data genome sequencing problem [86]. The authors address the issue of genome sequencing in big data by combining MapReduce frameworks and FPGA-based accelerators. Generally speaking, it can be divided into two major steps: 1) Assign tasks to all FPGAs through MapReduce stage; 2) Implement gene matching algorithm inside FPGA.

**2.2.3 Data Mining Accelerators**

Increasing amounts of data pose enormous challenges to data mining technologies and computer systems that adapt to these technologies. To provide flexibility, General-purpose CPUs and GPUs are not efficient at processing such algorithms. By contrast, hardware accelerators can provide high efficiency for some algorithms while satisfying their response time requirements.

For data mining algorithms, researchers at Georgia Institute of Technology put forward a simple machine learning accelerator framework [87], mainly focusing on the training phase of stochastic gradient descent algorithm. The framework regards the stochastic gradient descent algorithm as an intermediate layer and contains user-friendly APIs and declarations such as sum, pi, norm, sigmoid, log. Users programme through the given APIs and declarations. The compiler first translates the user-written program into a calculation graph in which each node represents a basic operation (+, -, *, /). Then according to the distance between the node and the destination node in the graph, the priority of the node is defined, and the corresponding scheduling result is generated according to the priority. Finally, the compiler will map each node in the graph to the pre-generated PE depending on the outcome of the scheduling.

Researchers at the University of Coimbra study Bayesian algorithms that implement random bitstreams in robotic applications using FPGA hardware [88]. For the hardware implementation of the Bayesian formula, the addition can be mapped to a simple AND gate, and the multiplication can be mapped to a simple multiplier. However, there is still a trade-off between prediction accuracy and computation time in robotic applications.

Experiments show that with a bitstream length of $10^6$ bits, the KL divergence could be lower than 0.0001, which is in the acceptable range. Moreover, at 25 MHz, it would require just 40 ms of computation time.

Researchers at the University of Rhode Island present a system that supports multiclass real-time SVM accelerators [89]. For a given model, we can know exactly how many cycles the accelerator needs to execute so as to support real-time row processing. Also because parameters in libsvm are stored in the form of floating-point numbers, adopting IEEE-754 floating-point calculation instead of the fixed-point calculation could provide higher accuracy and versatility. At the same time, a control signal is added for informing which kind of classification task the current task belongs to and which parameters are used so that different classification tasks can share the same hardware resources in a time-sharing manner. Besides, to obtain higher performance, the author also designs the computing unit into a pipeline mode so that each cycle could produce a result.

Researchers at Imperial College London propose a pulsatile binary tree model and implement a low-latency option pricing accelerator [90]. Experiments demonstrate that the exploit of 32bits fixed point could meet the accuracy requirement. In the binomial- tree option pricing model, each node's computation depends on its two child nodes, and most nodes have two parent nodes. Therefore, the pulse unit can be divided into two parts: 1) Calculation unit for calculating revenue; 2) Propagation unit, the current calculated value is passed to the neighbor unit for the revenue calculation of the next cycle of the parent node. Modern large-scale FPGA resources are enough to meet a binary tree model of 768 layers, due to which we can schedule PEs by layer. The first step is to calculate leaf nodes, and the second step is to calculate the parent nodes of leaf nodes, which could efficiently reduce latency and increase throughput.

**2.2.4 Graph and Database Accelerators**

For large-scale graph processing, the required data is random and irregular, and it is difficult for the processor's on-chip cache to ensure high data locality. Consequently, how to efficiently analyze, search and calculate large-scale graphs within an acceptable range of time has become a significant difficulty but hot spot in current research.

Reference [91] proposes a scalable graph processing PIM (Processing-in-Memory) accelerator, Tesseract, where computing units are placed in a 3D-stacked HMC. Each vault in the HMC is responsible for processing different sub-graph data and communicates via message passing between each other. Besides, to increase storage bandwidth utilization, two prefetching mechanisms suitable for graph data access patterns have been designed and implemented in Tesseract. Graphicionado [92] is a high-performance, energy-efficient graph processing accelerator that supports the mapping of different graph analytics applications to this accelerator framework. The work is based on GraphMat [93], a graph processing system previously proposed on a stand-alone platform. Graphicionado is a hardware implementation of the stand-alone graph processing system GraphMat. It profoundly analyzes the patterns that GraphMat adopts to access vertices and edges at various stages of graph processing to decrease the number of random access to vertices and edges. Reference [94] proposes an energy-efficient architecture for graph processing accelerators, which is mainly used for vertex centered graph algorithms for repeated iterations. Reference [95] proposes an FPGA-based high-throughput and energy-efficient graph processing accelerator, mainly characterized in that graph storage in external memory, edge-centered structure storage and optimization of the storage structure of graphs to make full use of memory bandwidth. Reference [96] mainly introduces accelerating SSSP (Single-Source Shortest Path) on FPGA, which mainly adopts the Bellman-Ford algorithm and stores graphs in external memory, and which could achieve the maximum data parallelism and process multiple edges per clock cycle. GraVF [97] is an FPGA-based vertex-centric distributed graph processing framework, in which designs of users could be automatically compiled into the target system. GraphOps [98] proposes a modular hardware dataflow library that could effectively lower the difficulty of building a graph calculation accelerator and shorten the hardware development time.

TuNao [99] designs a reconfigurable graph processing accelerator, whose main feature is exploiting the data locality of graph processing in the on-chip storage and providing reconfigurable function units to fulfill diversified operations in different graph processing tasks. Compared to the GPU, this accelerator achieves 1.58x and 25.56x better performance and energy efficiency. ForeGraph [100] implements a large-scale graph processing accelerator based on a multi-FPGA architecture. Its main feature is that the external memory of each FPGA board only stores one subgraph of the entire graph and the vertices and edges are loaded in turn into the FPGA for processing. This method effectively reduces the data interaction between the subgraphs and experimental results show that the average throughput of the accelerator more than 900 MTEPS. FPGP [101] is an FPGA-based graph processing accelerator that implements an improved vertex-centric graph processing framework on an FPGA. Meanwhile, FPGP makes a detailed theoretical analysis of related evaluation indexes such as performance speedup, energy efficiency ratio, power consumption ratio, and so on. From theoretical analysis and implementation results, it can be found that FPGP accelerators are better than the graphics computing systems of the single machine, with a good speedup and energy efficiency ratio. Reference[102] implements efficient key-value-pair access using a Cuckoo-based hash on the FPGA, which achieves 81.7% memory utilization through key-value-separated storage and efficiently reduces the delay of insertion, deletion and search operation by the pipeline and other means. The experimental results display that the speed of searching and deleting is 5x that of reference [103].

In recent years, applying hardware accelerators to database processing has once again become a research hotspot. As early as 2010, Rene Mueller et al. [104] have pointed out the potential of FPGAs in database applications and the challenges that FPGA database accelerators might face before becoming mainstream, such as how to integrate FPGA accelerators into database systems. Hashing is a rather critical and frequently-used operation in database applications. Consequently, FPGAs for accelerating hashing is widely used in web applications. Some researchers speed up hash operations in database applications by implementing entire hash tables on FPGAs [105, 106]. S. Richter et al. [107] expound the importance of choosing the correct hash function and mechanism in the database and demonstrate the necessity of a tradeoff between the robustness and performance of the hash function. However, after the release of heterogeneous multicore architectures such as Intel Xeon + FPGAs, Kaan Kara et al. [108] employ a heterogeneous multicore architecture of CPU-FPGA shared storage, successfully avoiding the overhead of data movement between FPGA accelerators and main memory. Evidently, this compromise has been broken, and the robustness and high-performance of hash operation could be achieved simultaneously. Similarly utilizing the shared memory structure of the FPGA and the CPU, David Sidler et al. [109] implement a reconfigurable hardware accelerator for two SQL clauses commonly used for string manipulation: LIKE and REGEXP_LIKE and integrate them into MonetDB. Compared with 10-core CPU, there is a significant performance improvement. Regular expression matching is also a typical operation in the database application. Thereby, utilizing FPGA to accelerate the regular expression matching operation is applied a lot in the field of the network. Zsolt István et al. [110] present an FPGA-based accelerator for regular expression processing, which could be configured at runtime to match specific regular expressions.

According to research statistics, about 40% of the data analysis work is done by SQL, and many other tasks can be overridden with SQL, and there are also various popular SQL systems such as HIVE, Spark SQL, Impala today.

Baidu research team proposes a Software-Defined Accelerator (SDA) [111] that accelerates various operations of SQL through hardware and abstracts the software interfaces for the upper layer use. This accelerator not only realizes a general acceleration scheme specific to big data processing also makes it easy to be integrated into distributed systems. Within the hardware framework, SDA consists of several PEs, each of which accomplishes one of five primary SQL operations (filter, join, sort, group by, aggregate). Such a structure is easy to scale and flexible, allowing to configure the type and number of PEs dynamically according to workload. Baidu researchers have implemented SDA through FPGA presently. The experimental results indicate that it could obtain 55x acceleration ratio in the actual query compared with the method of software implementation.

Traditional databases are inefficient when dealing with complex operations, and therefore, we can utilize hardware accelerators to expedite these complex operations. However, how to integrate hardware accelerators into a database system remains an open question. Researchers at Samsung [112] filter out extraneous data within the storage by integrating the ARM processor in an SSD to lower the amount of data sent to the CPU. Both platforms, IBM Netezza [113] and BlueDBM [114], deploy filters on data paths between storage/network and CPU to filter out irrelevant data in advance, thus improving database performance. The main work of Andreas Becher et al. [115] is to deploy a highly configurable hardware Bloom filter based on FPGA, which could significantly lessen the amount of data that needs to be processed. Focus on how to integrate an FPGA accelerator into a database engine from the system architecture level, Muhsen Owaida et al. [116] present Centaur which expresses the FPGA accelerator as a hardware thread and implements a flexible FPGA framework that allows hybrid parallel execution of database operations.

**2.2.5 High Performance Computing Accelerators**

With the advent of the big data era, people have an urgent need for efficient computing power. At the same time, with the continuous attention and deepening of research on reconfigurable computing accelerators in universities, reconfigurable computing accelerators have been applied to high-performance computing in succession and have attained excellent results.

Aiming at the common sorting problems in big data research, Wei Song and Dirk Koch et al. [117] conduct an acceleration for them by allowing pause and changeable sorting rate based on FPGA. This move significantly enhances the parallelization processing of tasks, thus increasing the speed of merging sequence and decreasing the overall sequencing time. Based on predecessor's research, Susumu Mashimo et al. [118] propose a new accelerator based on FPGA with optimized acceleration for sorting problems, and it possesses a higher frequency and larger throughput. Sang-Woo Jun et al. put forward a flash-based storage and FPGA-based acceleration solution [119] for the same big data sequencing problem, in which the accelerator includes highly efficient sorting networks and merge trees. Experiments reveal that this scheme is limited only by the flash storage bandwidth and achieves a good energy consumption ratio. Paul Grigoras et al. set out to address the problem of memory bandwidth limitations in FPGA-based Sparse Matrix-Vector Multiply (SpMV) implementations by utilizing non-zero lossless compression. By introducing a dictionary-based compression algorithm, they realized this idea on FPGA [120]. Gabriel Weisz et al. investigate and evaluate the performance of the currently widely used shared-memory processor-FPGA architectures [121] and carry out a discussion about FPGA acceleration problems in the architecture, which inspires the relevant research afterward. Considering the commonality of programming language between FPGA and GPU, Muhammed Al Kadi et al. [122] design FPGA-based architecture FGPU that could well support the language of GPU platform, such as OpenCL, which is of a certain significance for building GPU and FPGA computing platforms. Santhosh Kumar Rethinagiri et al. present two novel real-time heterogeneous platforms with three kinds of devices (CPU, GPU, FPGA), i.e., trigeneous platforms [123], for efficiently accelerating compute-intensive applications in both the high-performance computing and the embedded system domains. Compared with the current mainstream CPU-GPU, CPU-FPGA and quad-core CPU platform, it has a better energy consumption ratio.

Dennis Weller et al. exploit the OpenCL framework to optimize operations on the FPGA from the perspective of solving partial differential equations (PDEs) using FPGAs [124], and improve the performance of solving PDEs through FPGAs. Based on the idea of Near-data Processing (NDP), Mingyu Gao and Christos Kozyrakis analyze the shortcomings of previous architectures and present the Heterogeneous Reconfigurable Logic (HRL) for NDP systems [125]. HRL combines both coarse-grained and fine-grained logic blocks, separates routing networks for data and control signals, and uses specialized units to effectively support branch operations and irregular data layouts in analytics workloads. Compared to a single FPGA and a single CGRA, HRL has a higher performance. Burhan Khurshid and Roohie Naaz Mir combine Generalized Parallel Counters (GPCs) with FPGAs [126] and propose a heuristic algorithm that effectively maps the GPC logic onto the LUT structure on the FPGA. The algorithm could improve the computational efficiency by 33% -100% in most cases.

## 3 Trends and Challenges

Computer architecture researchers have noticed the impact of computer architecture on the performance of computer systems. In fact, never a particular computer architecture does exist, which can deliver optimal performance for all applications and tasks. For example, applications of deep learning possess a higher degree of parallelism and the most suitable architecture for handling such tasks is a multi-core parallel architecture. However, for some scientific-computing-related tasks that cannot be parallelized, the single-core architecture is the best-fit architecture, and it also needs to greatly promote the performance of the mononuclear. It is precisely because of the different computing tasks suitable for different computer architectures, so Gerald Estrin proposed the Reconfigurable Computing concept in the last century 60's. The reconfigurable computing includes a CPU as a central control unit and multitudinous reconfigurable processing units, where the reconfigurable computing units are controlled by CPUs and could be configured as an optimal architecture (ie, hardware programming) when performing corresponding tasks (eg, scientific computing, image recognition, pattern recognition, etc.). Reiner Hartenstein et al. [127] point out that reconfigurable computing makes the clock frequency of computing units much lower than the CPU, but the overall computing power is several times higher than the CPU, and power consumption is lower than the CPU.  Reconfigurable computing accelerators possess various advantages, but some drawbacks such as heavy reconfiguration overhead and high programming complexity also exist. This survey gives a summarization of the advantages and disadvantages of reconfigurable computing accelerators and looks ahead to the prospects and trends of reconfigurable computing.

## 3.1 Advantages of Reconfigurable Computing Accelerators

1. Low power consumption and high performance.

As mentioned earlier, low power consumption and high performance are the two most obvious advantages of reconfigurable computing accelerators. In terms of high performance, reference [128] demonstrates that the dot product operation of an elliptic curve cryptographic algorithm with a key length of 270 bits on a 66MHz FPGA chip, XC2V6000, only takes 0.36ms. On two 2.6GHz Intel Xeon computers, it takes 196.71ms to implement this algorithm with optimized software. It can be seen that reconfigurable computing accelerator has a performance 540x over the general-purpose CPU while the clock frequency is reduced by 40x.

2. Security.

With the advent of the big data era, data has played an increasingly important role. Correspondingly, as a carrier of data, the security of a computer has become crucial. At present, the first reaction of people to computer security concerns is that various antivirus software protects the computer. In fact, the software can only serve as a passive defender, unable to eliminate security risks. By contrast, the security could be better enhanced from the hardware architecture level, even eliminated.

3. Flexibility.

The inherent reconfigurability of reconfigurable computing accelerators makes them still valid for complex computing scenarios. For instance, multi-function hardware accelerators can handle frequent design changes well. Reconstructing accelerators with specific refactoring techniques could satisfy the changing needs of users. Hence, flexibility is also a highlight of reconfigurable computing accelerators. With the development of big data and cloud computing, the flexibility of reconfigurable computing accelerators is performed thoroughly. For instance, aiming to accelerate the reconfigurable computing ecosystem, Xilinx, the FPGA giant, released the reconfigurable acceleration stack in International Supercomputing Conference in mid-2016. Within the reconfigurable accelerator stack, Xilinx provides integration of the current popular application frameworks, including Caffe (Deep Learning Framework), FFMPEG (Image video processing), and SQL (Database). In this architecture, cloud server programmers can configure and apply the modules in the accelerator stack without using a hardware description language. Furthermore, Xilinx also offers a variety of libraries, in which FPGAs could be used more flexibly to accelerate by calling these libraries in the program.

4. Parallelism.

Practical experience has revealed that the pipeline can bring performance improvements and high parallelism could efficiently speed up the program execution. However, as the pipeline depth increases, it would result in complex structures, large hardware overheads and higher parallelism requirements for applications or programs themselves. If we rashly execute each program in parallel, a dramatic increase would be brought in system overhead and lose more than gain. As an increasing number of applications could be processed in parallel, the demand for parallelism will surge. Therefore, the study of parallelism in architecture enjoys a bright future.

5. Low Cost.

As chip manufacturing is progressively approaching nanotechnology, the superiority of FPGAs has become more apparent. In particular, by reconstructing multiple soft cores, multiple instruction set processors could be implemented on a single chip. According to the division of field computing tasks, different processor functions are implemented instantly, realizing that multiple functions are achieved by once chip design, thereby, drastically reducing the NRE (Non-Recurring Expenses) of designing. Through a vast market to assign costs, we will obtain the overall performance/price advantage while combining the flexibility of software implementation (general-purpose processor) with the high-performance benefits of hardware implementation (ASIC) [129].

## 3.2 Disadvantages of Reconfigurable Computing Accelerators

1. Reconfigurable Overhead Significantly

In the process of design and implementation of the reconfigurable computing accelerators, FPGAs generally demand to be configured, which includes synthesis, placement, and routing. However, depending on the accelerator complexity, these operations could cost tens of minutes or even hours. In line with the different timing of reconstruction, reconfiguration can be divided into static reconfiguration and dynamic reconfiguration. The former, also called compile-time reconfiguration, refers to that the reconfigurable hardware is configured one-time for one or several functions required by the system before the task starts, and these configured functions are forbidden to be changed during the execution of the whole task. Therefore, until the entire task completes, this reconfigurable hardware cannot be reconfigured to accomplish other tasks. The later, also called runtime reconfiguration, refers to that reconfigurable hardware is allowed to be configured at any time in the process of task execution. The runtime reconfiguration is mostly partial reconfiguration, usually adopting the context configuration mode. The key to implementing a dynamically reconfigurable system is to reconfigure the hardware as efficiently and fastly as possible. Also, if the configuration delay is too high, the computational continuity might be compromised, which will result in the overhead offsets the acceleration effect. For example, in the DISC II system, 25 to 71% of the execution time is spent on refactoring and 98.5% on the UCLAATR.

2. A Higher Programming Complexity

Although the concept of reconfigurable computing architecture has long been proposed, and there has been much more mature work, reconfigurable computing has not gained popularity before. There are two reasons for this: (1) The 40-year time from the appearance of reconfigurable computing to the early 21st century is the golden age of Moore's Law, during which technology updates every year and a half. So the performance improvements brought by the architectural updates are not as direct and forceful as technology updates; (2) With a mature system, traditional programming on the CPU adopts high-level abstract programming languages (such as Java, C / C ++.). However, reconfigurable computing requires hardware programming, generally using hardware programming Languages (Verilog, VHDL.) that would cost programmers much time to master.

## 3.3 Prospect

In the past several years of academic and industrial research, general-purpose CPU is most commonly used to handle different types of computing tasks, and with the ever-changing of upper applications, CPU frequency and performance are also continually rising. Therefore, for users' convenience, researchers are more inclined to choose the more mature CPU to handle different tasks. However, in recent years, with the failure of Moore's Law and the increasing demand for computer computing power by big data, the development of general-purpose processors has encountered bottlenecks. Blindly upgrading the frequency and performance of general-purpose processors can no longer meet the needs of users well. Therefore, domestic and foreign researchers have turned their attention to the design of dedicated accelerators in succession, aiming to design a dedicated accelerator suitable for different tasks to satisfy users' needs. In this way, CPU will be pulled away from the huge and complex work and only needs to distribute the computing task to the corresponding accelerator to complete.

Reconfigurable computing technologies have been widely used in many fields such as scientific computing, national defense and military affairs, aerospace and so on, and have achieved excellent results such as target matching, large numerical calculation, data mining, model simulation and other functions [130]. At present, the applications of reconfigurable computing technologies are also gradually expanding to civilian areas, and products have emerged in the fields of automotive electronics and network equipment. The research on reconfigurable computing technology is also flourishing in recent years. As can be seen from the theme of international conferences on reconfigurable computing technologies such as FPGA, FCCM, FPL, FPT, ReConFig, ERSA, RAW, ARC, research on reconfigurable computing technology has mainly focused on reconfigurable computing architecture, reconfigurable computing applications, reconfigurable computing tools, reconfigurable computing education, reconfigurable computing performance testing and so on.

With the rise of big data and artificial intelligence, the phenomenon that there are relatively more talents in the software but fewer talents in hardware has commonly existed in our country at present. The reason for this phenomenon is that to be proficient in the underlying hardware structure requires a full understanding and grasp of the composition and operation of the computer. Unfortunately, this is also a nodus in the computer field.

At the same time, as reconfigurable computing is precisely a computer research based on reconfigurable hardware platforms, which makes reconfigurable computing also become a major challenge in current research. By the comprehensive investigation of reconfigurable computing, we summarize the following three domestic research ideas that could play their expertise in the field of reconfigurable computing, for reference only.

(1) Research on reconfigurable computing is inseparable from reconfigurable computing professionals. However, at present, from the domestic situation, the training of talents in the field of reconfigurable computing is far from enough. Only a few domestic universities are equipped with related training modes and conditions. To catch up with or even surpass the relevant research level abroad in the field of reconfigurable computing, we need to establish a reasonable and perfect talent training mode in the area of reconfigurable computing and attract more young and energetic researchers or research teams to contribute to reconfigurable computing;

(2) As mentioned earlier, to obtain a long-term and robust development of reconfigurable computing-related research, we must have industry-related reconfigurable computing research platforms to support. However, the domestic work in this area is not enough yet. At present, the hardware devices and programming tools we use are mainly developed and manufactured by foreign developer giants such as Xilinx and Altera. Thus, we need to build our own powerful hardware devices and programming tools.

(3) Most research on reconfigurable computing has been currently based on specific applications such as accelerators based on neural networks, graph-based accelerators, and data mining algorithm accelerators. How to excavate more new type big data applications and design more targeted reconfigurable computing architectures to accelerate the applications will also be one of the hot topics in the future.

Generally speaking, both academia and industry have seen rapid advances in reconfigurable computing. Looking ahead, with high performance, high flexibility, low power consumption, low cost, high security and high parallelism, reconfigurable computing is bound to gain more excellent development and broader use.

## 4 Summary

With the rise of Artificial Intelligence and the arrival of big data era in recent years, data-intensive and compute-intensive applications have posed considerable challenges to computer processing ability. However, the lack of computer processing capacity restricts the development of these applications. Thus, people are turning more attention to the research of computer architecture, expecting to resolve conflicts from the hardware level. Surely, good expectations will bring pressure on the structural researchers as well as more opportunities and broader prospects for development.